# Half-filled intermediate bands Si material formed by energetically metastable interstitial sulfur atom

Ke-Fan Wang[a,b], Mingguo Liu[a], Yaping Ma[b], Zhenxiang Cheng[a,c,*], Yuanxu Wang[a, *], Xudong Xiao[b, *]

[a] Institute for Computational Materials Science, School of Physics and Electronics, Henan University, Kaifeng 475004, P. R. China

[b] Department of Physics, The Chinese University of Hong Kong, Shatin, Hong Kong, P. R. China

[c] Institute of Superconducting and Electronic Materials, University of Wollongong, NSW 2500, Australia

**Abstract**

Hyperdoped metastable sulfur atoms endow crystalline silicon with a strong sub-bandgap light absorption. In order to explore such metastable states, we develop a new high-throughput first-principles calculation method to search for all of the energetically metastable states for an interstitial sulfur atom inside crystalline silicon. Finally, we obtain sixty-three metastable interstitial states and they can be classified into ten types. Interestingly, twenty-eight (44% in total) of lower-energy metastable states can produce a well-isolated and half-filled intermediate band (IB) inside silicon forbidden gap, which makes sulfur hyperdoped silicon to be a desirable material for IB solar cells.

* E-mail: cheng@uow.edu.au (Prof. Z.X. Cheng), Tel.: +61-2-4221-1406;
* E- mail: wangyx@henu.edu.cn (Prof. Y. X. Wang), Tel./fax: +86-378-2388-1488;
* E-mail: xdxiao@cuhk.edu.hk (Prof. X.D. Xiao), Tel.: +852-3943-4388;

1. Introduction

Chalcogen atoms S, Se or Te-doped crystalline silicon has been studying for more than fifty years [1-3]. In those earlier studies, large efforts had been paid on the identification of impurity energy levels, local atomic structures, thermal diffusion of chalcogen atoms, and the related infrared detector working at low temperatures, etc [4]. At that time, chalcogen atoms were usually doped into silicon by thermal diffusion, and so the dopant density was below their saturated solid solubility ($\sim 10^{16} cm^{-3}$) inside silicon. In the last decades, however, this limit on doping density has been broken and it can be raised largely up to $\sim 10^{20} cm^{-3}$ (~1 at. %) by using ultrafast pulsed laser technologies [5, 6]. This optical hyperdoping of chalcogen atoms endows the crystalline silicon with a strong (>90%) and wide wavelength (250~2500 nm) light absorption [6-8], which make it be used for the fabrication of cheap room temperature infrared detectors [9-12].

However, annealing the chalcogen hyperdoped silicon even at low temperatures, such as 200ºC, begins to induce the attenuation of the sub-bandgap absorption [6,13]. Recently we model this attenuation from optically absorbing to unabsorbing state as a chemical decomposition reaction, whose activation energy was extracted to be about 0.34~0.47 eV, which energy is close to the calculated chalcogen-Si bond energies in high-energy interstitial sites [14]. Subsequently, we propose that the high-energy interstitial states, instead of the stable substitutional state, induce the strong sub-bandgap absorption inside chalcogen hyperdoped silicon. This opinion was also supported by the fact that temperature dependence of sub-bandgap absorption of S-, Se- or Te-hyperdoped silicon is same with that of their respective diffusion rates inside crystalline silicon through a interstitial mechanism [4, 15-16].

Till now, however, the knowledges about metastable interstitial states of a chalcogen atom inside crystalline silicon is still very limited [17-18]. In this paper, we develop a high-throughput first-principles calculation scheme which can search for all of the energetically metastable interstitial states for one sulfur atom inside crystalline silicon. Interestingly, near half of them own a well-isolated and half-filled IB which make sulfur hyperdoped Si to be a cheap and desirable IB semiconductor material.

## 2. Computational details

All of the first-principles calculations were carried out by density-functional theory (DFT) [19,20] method implemented plane-wave-based Vienna ab initio simulation package (VASP) [21-22]. The generalized gradient approximation (GGA) through the Perdew, Burke, and Ernzerhof (PBE) [23] functional was used to consider the exchange-correlation potential. Sulfur ion implantation plus nanosecond laser melting used by Kim et al. [6] can produce S-hyperdoped crystalline silicon samples without any detectable interior defects but also having a strong sub-bandgap absorption, which allows us to adopt a crystal model in the calculations. According to our previous analysis [14], we believe that the interstitial sites contribute mainly to the sub-bandgap absorption. Because the nanosecond laser melting is a thermal process [24], we suppose that the S atom preferentially occupies the energetically metastable interstitial site, which is defined as that position where if one sulfur atom locates at, the system total energy will be smaller than other system energies when it stays at the nearest neighbor positions.

To search for the energetically metastable interstitial sites, firstly we partition each *basis vector* of a silicon *primary cell* equally by 20 parallel planes and thus obtain a 21×21×21 lattice, producing 9261 grid points totally. The length of a *basis vector* is 3.84 Å and so the nearest distance between two neighbor grid points is 0.19 Å that is about 8.2% of a Si-Si bond length (2.35 Å). Then we construct a 3×3×3-$Si_2$ triclinic supercell (as shown in Fig. 1) centering the partitioned *primary cell* (denoted by nine red balls), and then let one sulfur atom locate at each grid point successively inside the central primary cell. The static energy of the supercell at each grid point was calculated and recorded. For these high throughput of static energy calculations, plane-wave energy cutoff of 300 eV is used; the k-point samplings are 3×3×3 according to the Γ-centered Monkhost-Pack (MP) method, which generates 6 k-points in the irreducible Brillouin zone (BZ); the tolerance for energy convergence is set to be $10^{-4}$ eV; Gaussian smearing with width of 0.05 eV is used for the electronic states near the Fermi level. In the end, we acquired 9261 static energies which were filtrated basing on the above definition on energetically metastable site. Finally, there are sixty-three fractional coordinates found and after removing the symmetry, ten types of metastable coordinates are remained.

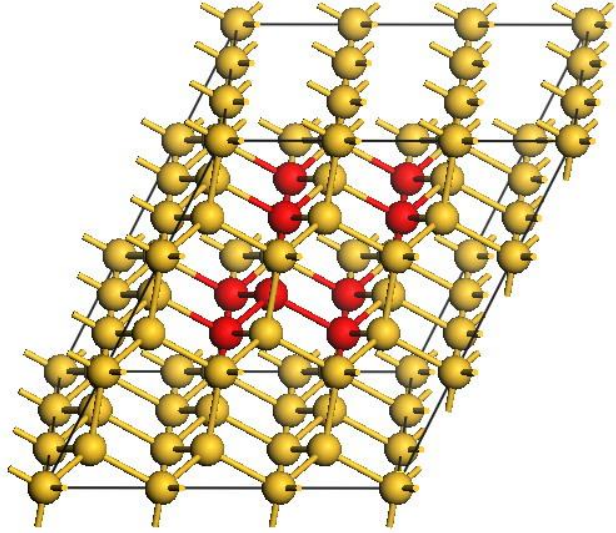

**Fig. 1.** A 3×3×3-$Si_2$ triclinic supercell. The nine red balls represent a central Si primary cell which is divided into 21×21×21 grid points and let one S atom locate at each grid point sequentially.

The real metastable site may not be nicely at the searched grid point, but around it closely with a distance less than 0.19 Å. In view of this possible case, we let the sulfur atom at each searched grid point relax a little while fixing all other silicon atoms in the supercell. Then this new obtained coordinate replaces the old one. Once the energetically metastable coordinates for an interstitial sulfur atom were determined, their relative positions in the central primary cell of the 3×3×3-$Si_2$ supercell (Fig. 1) are projected into the central primary cell of a new and larger triclinic 5×5×5-$Si_2$ supercell in order to reproduce a similar dilution level (~0.4 at. %) with that reported in experiments [6]. Then we perform a static calculation for this new 5×5×5-$Si_2$ supercell with one interstitial sulfur atom lying at one of the ten metastable coordinates. During these calculation, a larger plane-wave energy cutoff of 400 eV is used; the k-point samplings are 5×5×5 according to the Γ-centered MP method, which generates 39 k-points in the irreducible BZ; the tolerance for energy convergence is set to be $10^{-5}$ eV; Gaussian smearing with width of 0.05 eV is used for the electronic states near the Fermi level.

The GGA approach usually underestimates the semiconductor's band gap, and here we apply a scissor operator over the empty state of Si to recover the experimental band gap of 1.12 eV [25], which needs an energy shift of 0.54 eV. We do the same shift to the DOS of S-substitutional silicon and the searched ten types of S- interstitial Si. These calculation results are summarized in Table 1 and Fig. 2.

## 3. Results and discussion

All metastable structures studied here are not relaxed because that the interstitial S atom may move from one high energy metastable position to another low energy metastable position during structure relaxation. As the control samples, both crystalline Si and S-substitutional Si are also not relaxed and we use the experimental Si lattice constant (5.43Å) to construct the structures. As a result, the calculation results presented here are a little different with those after structure relaxation. For example, as shown in Table 1, energy differences between valence band (VB) and IB ($\Delta E_{VI}$), conduct band (CB) and IB ($\Delta E_{IC}$), as well as the energy width of IB ($\Delta E_{IB}$) for the unrelaxed S-substitutional Si are respectively 0.31 eV, 0.54 eV, and 0.27 eV, and those corresponding values for relaxed structure are 0.32 eV, 0.52 eV, and 0.28 eV. The results of relaxed S-substitutional Si agrees well with those reported previously by others [25].

**Table 1.** Structural, energetics, and electronic features of the referential crystalline Si and S-substitutional Si, as well as the searched ten types of metastable structures for an interstitial S atom inside crystalline silicon. Here, all structures are not relaxed except that the relaxed crystalline Si and relaxed S-substitutional Si are shown here for reference. Column 1 shows the structure's name and its degeneracy; Column 2 is the fractional coordinate of the interstitial S atom inside the 5×5×5-$Si_2$ supercell; Column 3 is the bond length and coordination number with S atom (cutoff distance for S-Si bonding is 2.7 Å); Column 4 is the structure's formation energy assuming the reservoirs to be unrelaxed bulk Si and isolated S atom; Column 5, 7 and 6 show the energy difference between VB and IB ($\Delta E_{VI}$), IB and CB ($\Delta E_{IC}$), as well as IB bandwidth ($\Delta E_{IB}$), respectively; Column 8 presents whether the IB is partially filled or not.

| Structures and degeneracy | Fractional coordinate of S atom | Bond length (Å) and numbers | Formation energy per S (eV) | $\Delta E_{VI}$ (eV) | $\Delta E_{IB}$ (eV) | $\Delta E_{IC}$ (eV) | Partially filled? |
|---|---|---|---|---|---|---|---|
| c-Si | - | 2.35 | - | - | - | - | - |
| Ss | - | 2.35, 4 | -2.62 | 0.31 | 0.27 | 0.54 | no |
| c-Si-relax | - | 2.37 | - | - | - | - | - |
| Ss-relax | - | 2.47, 4 | -2.81 | 0.32 | 0.28 | 0.52 | no |
| S1, 3 | 0.4, 0.44, 0.44 | 1.18, 1; 1.33, 1; 2.70, 1 | 52.88 | - | - | - | - |
| S2, 6 | 0.45, 0.4, 0.43 | 1.2, 1; 1.34, 1; 2.64, 1 | 49.45 | - | - | - | - |
| S3, 6 | 0.4, 0.4, 0.49 | 1.37, 1; 1.73, 1; 2.11, 1 | 17.51 | ~0 | 0.12 | 0.88 | no |
| S4, 12 | 0.514, 0.492, 0.48 | 2.15, 1; 2.18, 1; 2.35, 1; 2.50, 1; 2.65, 1; 2.67, 2 | 0.45 | ~0 | 0.24 | 0.83 | yes |
| S5, 2 | 0.5, 0.5, 0.5 | 2.35, 4 | 0.44 | ~0 | 0.07 | 0.97 | yes |
| S6, 6 | 0.54, 0.55, | 2.22, 2; | 0.39 | ~0 | 0.20 | 0.87 | yes |

|  | 0.57 | 2.36, 1; 2.46, 1; 2.58, 2; 2.68, 1; | | | | | |
|---|---|---|---|---|---|---|---|
| S7, 6 | 0.58, 0.54, 0.54 | 2.24, 3; 2.47, 3; | 0.31 | 0.11 | 0.18 | 0.79 | yes |
| S8, 1 | 0.537, 0.537, 0.537 | 2.22, 3; 2.42, 3; | 0.29 | 0.18 | 0.18 | 0.73 | yes |
| S9, 14 | 0.52, 0.52, 0.44 | 2.22, 3; 2.31, 3; | 0.27 | 0.33 | 0.18 | 0.59 | yes |
| S10, 7 | 0.516, 0.516, 0.516 | 2.22, 3; 2.36, 3; | 0.27 | 0.26 | 0.18 | 0.66 | yes |

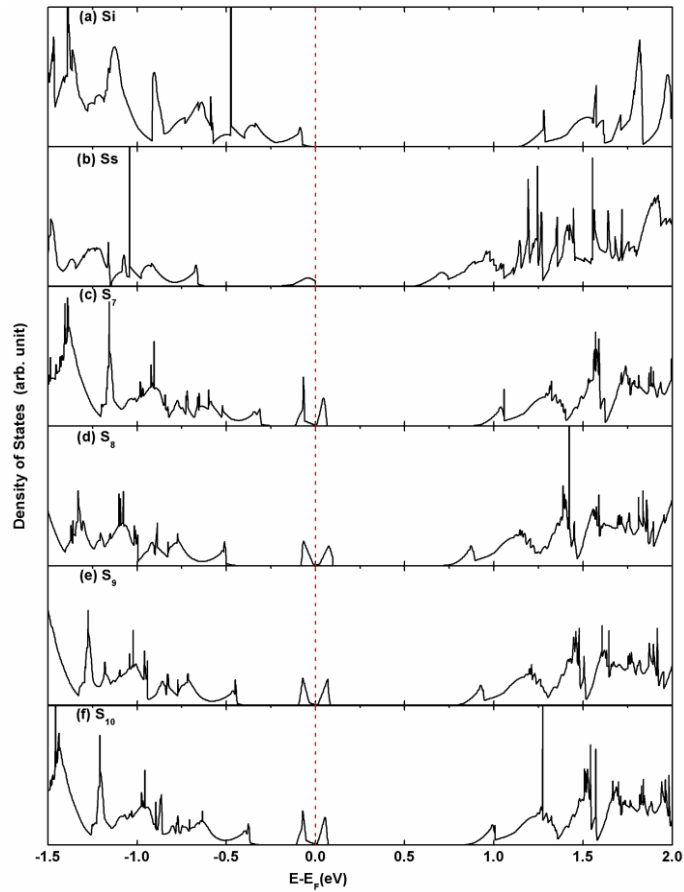

**Fig. 2.** Density of states (DOS) of crystalline Si (a), S-substitutional Si (b), and four types of low-energy metastable interstitial structures with half-filled IBs, including $S_7$ (c), $S_8$ (d), $S_9$ (e) and $S_{10}$ (f).

In Table 1, the metastable structures S1 and S2 both need very high formation energies, 52.88 eV and 49.45 eV per S atom, respectively, because that both of them have two much shorter Si-S bonds (1.18-1.2 Å and 1.33-1.34 Å) than the normal relaxed Si-S bond in substitutional site (2.47 Å). Moreover, S1 and S2 both do not form any IBs inside the Si forbidden gap. As the Si-S bond length and coordination number increase, the searched metastable structures S2, S3, S4, S5 and S6 needs less and less formation energy, namely, from 17.51 eV decreasing continuously down to 0.39 eV. They all form an IB inside the Si forbidden gap, but these IBs almost overlap with the VBs, and so they are not suitable IB materials. For an IB solar cell, the absorber's IB must be isolated from the VB and CB, otherwise a fast deexcitation will happen to the photo-generated electron in IB and CB as a result of interaction with phonons.

But for the four types of low-energy metastable structures S7, S8, S9 and S10, they all form a well-isolated IB which has an energy difference of 0.11-0.33 eV from VB and 0.59-0.79 eV from CB, and interestingly, the IBs are all half-filled by electrons, as shown in Fig. 2(c)-(f). This half-filled IB is another crucial characteristics for an IB semiconductor material to realize its very high theoretical conversion efficiency because that half-filled IB can act as a very efficient stepping-stone for *simultaneous* light-electron transition from VB to IB and IB to CB, which process can efficiently pump an electron from VB to CB by two low energy (smaller than the bandgap $E_g$) photons [26]. However, the filled IB of S-substitutional Si (Fig. 2(b)) have to permit the transition from IB to CB firstly and can the generated hole in IB allow the electron jumping from VB to IB then. Si the light-electron transition efficiency is very low in comparison to that of the half-filled IB.

Sánchez et al [25] found that chalcogen (S, Se or Te) substitutional doping can introduce an filled IB into silicon forbidden gap, and in order to realize the hall-filled IB, they suggested the additional co-doping of group III elements (such as B or Al). Our calculated DOS in Fig. 1(c)-(f) show that single S doping can realize the half-filled IB if the energy of nanosecond pulsed laser is optimized carefully (such as 0.27-0.31 eV calculate here), which will make the usable Si IB material more easily available. In view of the present dominance of Si solar cells, this convenient method to produce Si IB material potentially has a large technical importance.

To explore the underlying mechanism accounting for the formation of half-filled and well-isolated IB, we calculated the partial DOS (PDOS) of metastable structures S4, S7 and S9, as shown in Fig. 3. From the PDOS of S4 (Fig. 3(a)), we can see that its IB mainly originates from S $3p$ orbitals hybridized with Si $3p$ orbitals, and the strongest hybridizing happens at the atomic distance of 2.35 Å between the S atom and Si127 atom. The IB's filled orbitals mainly consist of S $3p$, Si127 $3p$, Si137 $3p$, and Si135 $3p$, while its empty orbitals mainly contain S $3p$, Si127 $3p$, Si128 $3p$, Si126 $3p$, and Si137 $3s$. For these orbitals, there are totally sixteen electrons and still needs another twenty-two electrons to be completely filled up, so the S4's IB is nearly half-filled. For the metastable structures S7, a strong coupling happens between one S 3p orbital and three Si 3p orbitals. There are ten electrons in these orbitals already, which need another fourteen electrons to fill their empty orbitals, so the formed IB is also almost half-filled. Similar situation happens to the metastable structure S9.

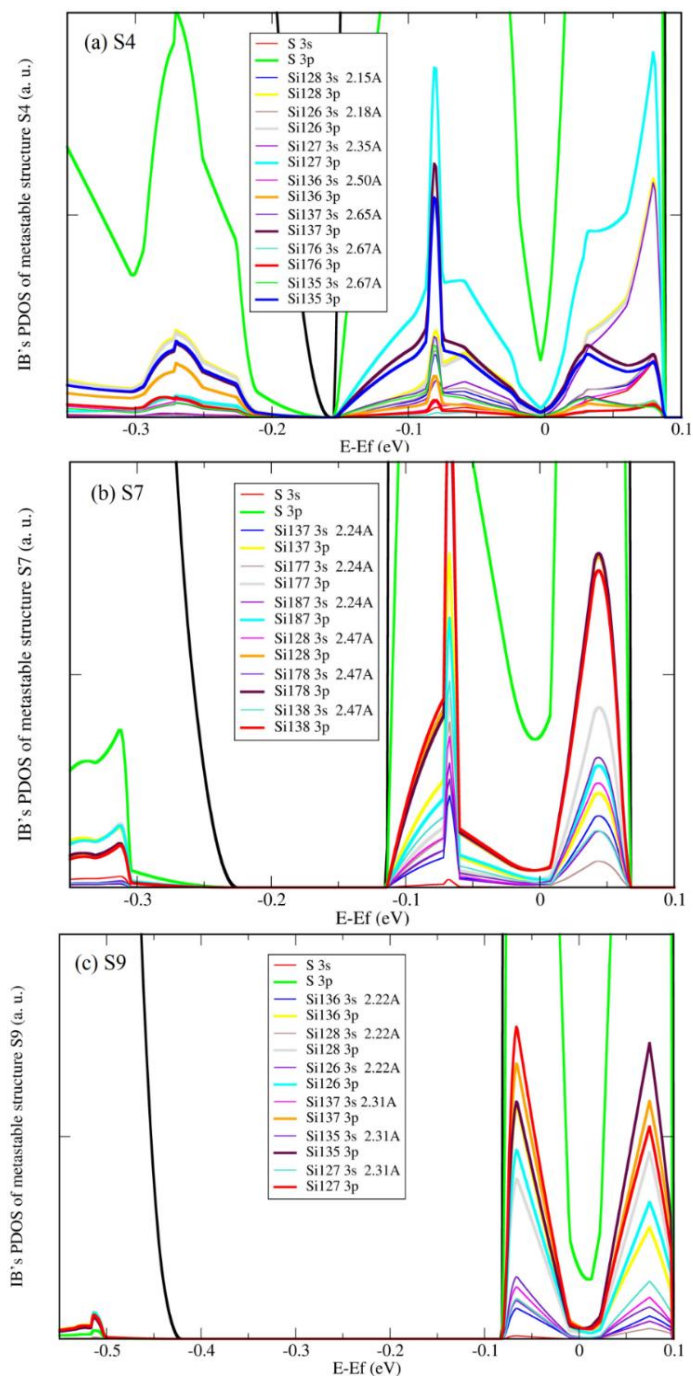

**Fig. 3**. Partial DOS of three typical metastable structures S4, S7 and S9

It is noteworthy that the S4's IB does not separate obviously (~0 eV) from its VB, while the IBs in S7 and S9 are well-isolated to their VB, with an energy difference of

0.11 eV and 0.33 eV, respectively. The VB maximum (VBM) of S4 and S7 comes from the filled S 3$p$ orbital, but that of S9 is mostly from the coordinated Si 3$p$ orbitals. And the contribution of S 3$p$ orbital to VBM decrease continuously from S4, S7 to S9. This decrease is probably due to the enhanced hybridizing between S 3$p$ orbital and the Si 3$p$ orbitals, and more electrons delocalize from S 3$p$ orbital to Si 3$p$ orbitals. The new formed hybridizing orbitals have a larger energy difference from the S 3p orbitals of IB, and so the VB and IB are well-isolated.

## 4. Conclusion

In summary, we develop a new high-throughput first-principles calculation method which can search for all of the metastable interstitial states for one S atom inside crystalline silicon. This method can also be extended to more general case that one doping atom locates inside a crystalline sample. Finally, sixty-three of metastable interstitial structures are found and twenty-eight of them own a well-isolated and half-filled IB inside Si forbidden gap, which suggests a much easier technique to obtain usable silicon IB material.

**Acknowledgments**

This work was supported by the National Natural Science Foundation of China (Grant Nos. 61204002, 11305046 and U1404619).